%% file: main.tex
\documentclass[10pt]{IEEEtran}
\usepackage{setspace}
\usepackage{array}
\usepackage[super]{nth}
\usepackage{verbatim}
\usepackage{cite}
\usepackage{enumitem}
\usepackage{footnote}
\usepackage{graphicx}
\usepackage{longtable}
\usepackage[table,xcdraw]{xcolor}
\usepackage{adjustbox}
\usepackage{tabularx}
\usepackage[T1]{fontenc}
\usepackage{amsmath}
\usepackage{array}
\usepackage{comment}
\usepackage{soul}
\usepackage{amssymb}
\usepackage{mathdots}
\usepackage{float}
\usepackage{subcaption}
\usepackage[font=scriptsize,labelfont=bf]{caption}
\usepackage{url}

\usepackage{mathptmx}
\usepackage[papersize={8.5in,11in,top=0.75in, bottom=0.75in, left=0.75in, right=0.75in}]{geometry}

\def\BibTeX{{\rm B\kern-.05em{\sc i\kern-.025em b}\kern-.08em
    T\kern-.1667em\lower.7ex\hbox{E}\kern-.125emX}}

\begin{document}

\title{Approximating Aggregated SQL Queries With LSTM Networks}
 \author{\IEEEauthorblockN{Nir Regev, Lior Rokach, Asaf Shabtai}
 \IEEEauthorblockA{Dept. of Software and Information Systems Engineering\\
 Ben-Gurion University of the Negev\\
 Be'er Sheva, Israel\\
 Email: nirre@post.bgu.ac.il, \{liorrk,shabtaia\}@bgu.ac.il}
}

\maketitle
\begin{abstract}
Despite continuous investments in data technologies, the latency of querying data still poses a significant challenge.
Modern analytic solutions require near real-time responsiveness both to make them interactive and to support automated processing. 
Current technologies (Hadoop, Spark, Dataflow) scan the dataset to execute queries. 
They focus on providing a scalable data storage to maximize task execution speed. 
We argue that these solutions fail to offer an adequate level of interactivity since they depend on continual access to data. 
In this paper we present a method for query approximation, also known as approximate query processing (AQP), that reduce the need to scan data during inference (query calculation), thus enabling a rapid query processing tool.
We use LSTM network to learn the relationship between queries and their results, and to provide a rapid inference layer for predicting query results. 
Our method (referred as ``Hunch``) produces a lightweight LSTM network which provides a high query throughput. 
We evaluated our method using twelve datasets and compared to state-of-the-art AQP engines (VerdictDB, BlinkDB) from query latency, model weight and accuracy perspectives.
The results show that our method predicted queries' results with a normalized root mean squared error (NRMSE) ranging from approximately 1\% to 4\% which in the majority of our data sets was better then the compared benchmarks. 
Moreover, our method was able to predict up to 120,000 queries in a second (streamed together), and with a single query latency of no more than 2ms.

\end{abstract}

\input{Sections/introduction.tex}

\input{Sections/relatedworks.tex}

\input{Sections/method.tex}

\input{Sections/eval.tex}

\input{Sections/discussion.tex}

\bibliographystyle{IEEEtran}
\bibliography{Bibliography}
\end{document}

%% file: Sections/introduction.tex
\section{\label{sec:intro}Introduction}
While Big Data opens new possibilities for extracting unprecedented insights, this may come at the price of high processing latency and increased computational resource requirements for answering queries over large data sets~\cite{chaudhuri2017approximate}. 
AQP can be particularly useful for data analysts, who often need to discover and explore large new data sets.
This task requires a fast, efficient, and cost-effective query engine and does not necessarily rely on exact answers. 
In addition, when data becomes too large to fit in a single machine, data processing platform vendors (e.g., Hadoop, Spark, Google Cloud Dataflow) address this challenge by scaling out resources. 
However, this strategy may be increasingly cost-prohibitive and could be inefficient for large and distributed data sources~\cite{sivarajah2017critical}. 
As a result, it has been shown that data exploration could be successfully performed in an approximate fashion~\cite{slkezak2018new}. 

In this research we introduce a new method that can produce high-value approximated results for SQL queries by training a LSTM network, without a-priori domain knowledge, to learn the relationship between the different elements of SQL queries and their results.
Our method processes queries on large datasets rapidly and in a fixed response time, regardless of the data size.

We applied the proposed method on twelve data sets taken from the technology industry. 
We evaluated the method predictions (approximations) using large hold-out testing sets of queries. 
Our results show that the proposed solution approximates query results within a controlled range of normalized error (NRMSE), between 1\% to 4\%. 
In terms of execution latency, query latency (QL) ranges from approximately 2 ms/q (millisecond per query) to 30 ms/q. 
We also evaluated method performance on large batches of queries (processed in parallel on a GPU). 
Our method demonstrated a query throughput (QT) of approximately 2,000 queries per second to 120,000 queries per second, depending on input dimension and the LSTM network architecture (number of neurons and number of hidden layers). 

In summary, the contributions of this paper are as follows:

\begin{itemize}
    \item we introduce a novel approach for producing lightweight data representation layer in the form of a NN;
    \item we propose an effective query processing method, with lightning-fast query response times for big data platforms;
    \item we present a forceful concurrent approach to process SQL queries with GPU technology;
    \item Finally, we make our code and datasets publicly available @\url{ https://github.com/nirre1401/aqp.git}
\end{itemize}

%% file: Sections/relatedworks.tex
\section{\label{sec:relatedworks}Related Works} 

\textbf{Query approximation by sampling.} Previously, different approaches to approximate database queries have been introduced, with the majority based on executing queries over smart samples of data~\cite{mozafari2015handbook}.
These approaches rely on the ability to use a statistical method to deliver an approximated result within confidence interval. 
While several research projects have explored the benefits in data sampling~\cite{bagchi2007deterministic},~\cite{babcock2001sampling},~\cite{chuang2009feature}, these methods are not widely used in streaming engines~\cite{chandramouli2014trill,zaharia2013discretized} with the noteworthy exception found in the SnappyData project~\cite{ramnarayan2016snappydata}, which uses the notion of High-level Accuracy Contract (HAC), which is also used in VerdictDB ~\cite{park2018verdictdb}.
When memory is limited (as is often the case), sampling may help to enable in-memory processing.

\noindent \textbf{Interactive approximate query processing.} The SnappyData~\cite{ramnarayan2016snappydata} engine, developed in 2015, was designed to support query approximation in streaming, transnational, and interactive systems. 
It is based on many insights gained from the BlinkDB project~\cite{agarwal2013blinkdb}. 
Spark, a contemporary distributed in-memory data processing engine, manages smart query caching (referred to as delta update queries) with confidence intervals utilized to minimize loss. 
It also handles on-line aggregation to reduce processing latency by presenting preliminary approximated results immediately on processing a small portion of the whole dataset~\cite{zeng2015g}. 
Oracle Database uses the HyperLogLog (HLL) algorithm for approximate `count distinct' operations. 
\\
\textbf{Query approximation with ML.}
As mentioned above, the database research community has proposed novel techniques for AQP that could give approximate query results in orders of magnitude faster
than the time needed to calculate exact results. In this
work, the usage of deep generative model, specifically Variational auto-encoder (VAE), for answering
aggregate queries specifically for interactive applications such as data exploration and visualization\cite{thirumuruganathan2020approximate}. 
Similar to our approach, this work utilized ML models to approximate aggregated SQL queries  \cite{Fotis2020ArXiv}. Specifically, gradient Boosting Machines (GBM), XGBoost and LightGBM were trained to predict the aggregated queries' result.

\begin{table*}[h]
\tabcolsep=0.008cm
\small
\centering
\caption{AQP project benchmark comparison table.}
\label{tab:relatedworks}
\begin{tabular}{|c|c|c|c|c|c|c|c|c|} 
\hline
Paper                                                                                & Name                                                                       & \begin{tabular}[c]{@{}c@{}}Flat Query Latency\\in sec.\\(per 1Tb data)\end{tabular} & \begin{tabular}[c]{@{}c@{}}Guaranteed \\Error bound\end{tabular} & \begin{tabular}[c]{@{}c@{}}GPU~\\Support\end{tabular} & \begin{tabular}[c]{@{}c@{}}Training~\\Requirement\end{tabular} & \begin{tabular}[c]{@{}c@{}}Preprocessing\\/Sampling\\Requirement\end{tabular} & \begin{tabular}[c]{@{}c@{}}Queries Batch\\Concurrent \\Processing\\Support\end{tabular} & \begin{tabular}[c]{@{}c@{}}Result\\Confidence\end{tabular}  \\ 
\hline
\cite{Agarwal2012PVLDB}\cite{dokeroglu2014improving}~       & Hive Hadoop                                                                & 400                                                                                 & no                                                               & yes                                                   & no                                                             & yes                                                                           & yes                                                                                     & NA                                                          \\ 
\hline
\cite{Todor2016}\cite{Agarwal2012PVLDB}            & Hive Spark                                                                 & 40                                                                                  & no                                                               & yes                                                   & no                                                             & yes                                                                           & yes                                                                                     & NA                                                          \\ 
\hline
\cite{agarwal2014knowing}\cite{agarwal2013blinkdb} & BlinkDB                                                                    & 2                                                                                   & 2-10\%                                                           & no                                                    & no                                                             & yes                                                                           & yes                                                                                     & 95\%                                                        \\ 
\hline
\cite{ramnarayan2016snappydata}\cite{mozafari2017snappydata}~  & SnappyData                                                                 & 1.5                                                                                 & NA                                                               & no                                                    & no                                                             & yes                                                                           & yes                                                                                     & NA                                                          \\ 
\hline
\cite{he2018demonstration}\cite{park2018verdictdb}   & VerdictDB                                                                  & 1                                                                                   & 2.6\%                                                            & no                                                    & no                                                             & yes                                                                           & yes                                                                                     & 95\%                                                        \\ 
\hline
\cite{jayachandran2014combining}\cite{Kamat2014IEEE}           & DICE                                                                       & 0.5                                                                                 & 10\%                                                             & no                                                    & no                                                             & yes                                                                           & no                                                                                      & NA                                                          \\ 
\hline
\cite{thirumuruganathan2020approximate}~                                           & DeepGen                                                                    & NA                                                                                  & 0.1-1.25\%                                                       & yes                                                   & yes                                                            & yes                                                                           & yes                                                                                     & No                                                          \\ 
\hline
\cite{Fotis2020ArXiv}~                                              & ML AQP                                                                     & 20                                                                                  & 1-5\%                                                            & no                                                    & yes                                                            & yes                                                                           & no                                                                                      & NA                                                          \\ 
\hline
                                                                  \textbf{Our method}                   & \begin{tabular}[c]{@{}c@{}} \textbf{Hunch (DL)}\\\textbf{AQP}\end{tabular} & \textbf{10}                                                                         & \textbf{\textless{}2.5\%}                                        & \textbf{yes}                                          & \textbf{yes}                                                   & \textbf{yes}                                                                  & \textbf{yes}                                                                            & \textbf{NA}                                                 \\
\hline
\end{tabular}
\end{table*}

%% file: Sections/method.tex
\section{\label{sec:method}Proposed Method}
The proposed method uses a process to generate the training set which is used for fitting the query approximation model. 
This process is divided into four phases: (1) generating artificial SQL queries (Section~\ref{sec:querygen}), (2) obtaining the labels for the training set by executing the queries on the database (Section~\ref{subsec:execute_queries}), (3) inducing an encoder for transforming the queries into numeric matrices (Section~\ref{subsec:Encoding_queries}) and finally (4) training a neural network to approximate the queries labels (Section~\ref{subsec:model_fit}). 




\subsection{\label{sec:notations}Query template and notations}
As an example, assume a table `transactions' that includes computer sales from either physical or online shops. 
The table includes the attributes: `hour' (time of transaction), the `store\_type' (physical or online), the `computer\_type', the `harddisk\_size' (hard disk size in Gb), and the `sales' (amount in \$).
The method is designed to generate many instances of queries conforming to a query template defined by the following:

\begin{itemize}

\item $attr^{(c)}$ -- denotes a continuous data attribute in the dataset (e.g., `harddisk\_size').

\item $attr^{(n)}$ -- denotes a nominal data attribute in the dataset (e.g., `computer\_type', `store\_type').



\item $A=\{a_1, a_2, \ ...\}$ -- denotes the set of optional aggregation functions (e.g., avg, count).

\item $a_i(attr)$ -- denotes an aggregation function $a_i \in A$ that is applied on valid attribute $attr$ (either $attr^{(c)}$ or $attr^{(n)}$) in a select query (e.g., avg(`sales'), max(`revenue'), or count(`id')).

\item $between_{attr^{(c)}}(l,u)$ - a `between' constraint argument defined on a continuous data attribute $attr^{(c)}$, where $l$ is a lower bound and $u$ is an upper bound on the values of $attr^{(c)}$. 
\item $in_{attr^{(n)}}(v_k)$ - an `in' constraint argument defined on a nominal data attribute $attr^{(n)}$, where  $(v_k)$ is a single possible member of $attr^{(n)}$. 
\end{itemize}


\subsection{Support ``Group By" queries with multiple aggregations}
We support queries with "Group By" clause on multiple $attr^{(n)}$.
This enables flexibility in exploring and analyzing large datasets on one side, but poses the following two challenges on the other side: (1) learning different data distribution (characterized by the aggregation functions), and (2) learning an output which may have varying dimensions (while NN models expects fixed output types and dimension). 
The latter results from the fact that a "Group By" query can return a table of one or more rows as shown by the example in Table~\ref{table:groupbyresultset} which is the result of the following running example query: \noindent 

\begin{small}
\begin{verbatim}
SELECT computer_type, store_type,
AVG(sales), MEDIAN(revenue) 
FROM transactions WHERE 
hour between (20 and 23) AND 
harddisk_size between (121 and 820)
GROUP BY computer_type, store_type
\end{verbatim}
\end{small}

To tackle these challenges, we transform each "Group By" query to multiple `flat' (with no "Group By" term) queries with a single aggregation function. These, by definition, returns one scalar. This way, every LSTM network has an output layer which consists of a single linear output that is trained to learn a specific aggregation function distribution.



\begin{table}[h]
\tabcolsep=0.006cm
\small
\caption{An example for a group by result set.}
\label{table:groupbyresultset}
\begin{tabular}{|c|c|c|c|}
\hline
\textbf{store\_type} & \textbf{computer\_type} & \textbf{AVG(sales)} & \textbf{MEDIAN(revenue)} \\ \hline
online               & MAC                     & 102                 & 85                       \\ \hline
online               & IBM                     & 80                  & 82                       \\ \hline
phisical             & MAC                     & 95                  & 61                       \\ \hline
phisical             & IBM                     & 94                  & 50                       \\ \hline
\end{tabular}
\end{table}

\subsection{\label{sec:querygen}SQL queries generation}
The goal of this phase is to generate a large representative set of aggregated SQL queries $q_i$ that broadly represents the raw data.

\noindent \textbf{Define query template parameters.}
A query template is define by the domain expert which provides for each query template: (1) the SELECT clause parameters, and (2) the filter template (i.e., the WHERE clause parameters).

In this phase, first, the domain expert choose a set of aggregation functions and a set of data attributes. 
Then, the method constructs a SELECT clause consisting of the selected aggregation functions, which are applied on a set of valid data attributes, either continuous or nominal, $\{a_i(attr_j)\}$. \\
All aggregation functions can be applied on continuous data attributes, whereas the only aggregation functions that can be applied on a nominal attribute are ${'count'}$ and ${'countDistinct'}$.
In our example, assuming the domain expert chooses to apply all aggregation functions $A$ on all valid $\{a_i(attr_j)\}$, this results in the select clause:

\begin{small}
\begin{verbatim}
SELECT AVG(sales), MEDIAN(sales),
AVG(revenue), MEDIAN(revenue)
\end{verbatim}
\end{small}

As mentioned, each $\{a_i(attr_j)\}$ will have a designated model (see Section \ref{subsec:model_fit}) fitted to learn its distribution.
This means that the training set will be split for each $\{a_i(attr_j)\}$ and learned separately. 
In our example the first training set will consist of queries with AVG(sales) in the select clause, the second training set will consist of queries with MEDIAN(sales) and so on.

Next, the domain expert can choose a filter template which includes the list of continuous data attributes $attr^{(c)}$ and nominal data attributes $attr^{(n)}$ that can be included in each query.
Then, for each $query template$ defined by the domain expert, the method generates a set of $query instances$ as follows.

\noindent \textbf{Generating filter.} In this step, the method generates a rich set of filters applied on (1) continuous data attributes $attr^{(c)}$ and (2) nominal data attributes $attr^{(n)}$ in the following manner.

\begin{enumerate}
\item For each $attr^{(c)}$, we calculate the intervals defined by: the minimum value, the first quartile (25\%), the median, the third quartile (75\%), and maximum value (four intervals).  
In order to select the lower and upper bounds of a continuous attribute constraint $between_{attr^{(c)}}(l,u)$ we select two intervals randomly.
Then, from each selected interval, we randomly choose a value (from a uniform distribution). 
This results into two numeric values which form a filter 
, such that the smaller value will define the lower bound and the larger value will define the upper bound.
For instance, assume the $harddisk\_size$ continuous attribute values spanning from 1 to 1000.
Given those values, minimum=0, 25\% quartile=250, median=500, the 75\% quartile=750 and maximum=1000, and assuming the selected intervals are [0,250] and [750,1000] the continuous constraint might take the values $between_{harddisk\_size}(121,820)$

\item To construct a nominal filter, the method uses an "IN" constraint argument defined on a nominal data attribute $attr^{(n)}$, filtered by $v_k$ - a possible member of $attr^{(n)}$. 
To determine which member to use in each filter, the method constructs a "Group By" term on the nominal attribute and once the query is executed against the dataset, the method systematically extracts all possible combinations of members which exists in the result set and constructs a nominal filter for each one.
In our example, one combination of members for $store\_type$ and $computer\_type$:

$\{in_{store\_type}('online'), in_{computer\_type}('Mac') \}$


\item Finally, each of these combinations of nominal filters is paired with each of the continuous filters to form a query filter; for example,
$\{between_{harddisk\_size}(121,820),
\\
in_{store\_type}(`online'), in_{computer\_type}(`Mac') \}$.

\end{enumerate}

\subsection{\label{subsec:execute_queries}Obtaining labels for the training set}
To build a supervised dataset for training the proposed method needs to obtain real query results; therefore, our method runs the set of generated queries $Q$ against the data source. 
As LSTM model requires a relatively large number of training examples, the system needs to generate and execute hundreds of thousands of queries. 
However, by using "Group By" queries, the actual size of $Q$ is much smaller.

\subsection{\label{subsec:Encoding_queries}Encoding queries}

\begin{figure*}[h]
\centering
\includegraphics[width=\linewidth]{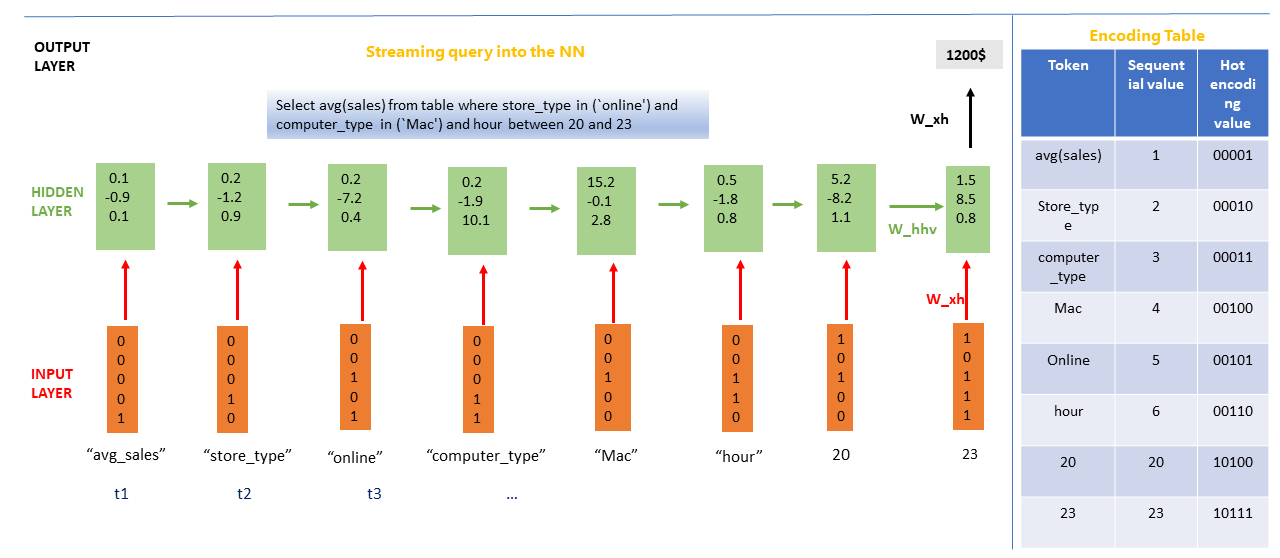}
\caption{Query encoding process: for each query token, the token encoding is streamed into the LSTM model.}
\label{fig:encoding_method}
\end{figure*}

At this stage, a list of "flat" SQL queries and their real label (result) is available. 
Since neural networks can take only numeric input, we encode the queries into numeric matrices (see Figure~\ref{fig:encoding_method}) via an encoder model which is constructed on the fly (during SQL queries generation), making use of a multi-hot encoding technique. 
The reason for choosing this type of encoding is to enable a distinctive representation for discrete data entities. 
That way, entities like `Mac' and `IBM' will be represented by perpendicular bit-wise vectors to signal the LSTM which entity exists in the input (query).
The encoding process starts by mapping all unique query tokens that exist in the training set $Q$ and assigns each with a numeric sequential value, as illustrated in Figure~\ref{fig:encoding_method} (for example, the token avg(sales) is mapped to the value 00001).
Each numeric value is then transformed into binary (base 2) numeric representation.
Numeric query tokens (scalars) are also transformed into their binary (base 2) representation. 

\subsection{\label{subsec:model_fit}LSTM network generation and training}
At this point, the algorithm pipeline builds an LSTM network. 
The network weights were initialized according to the Xavier initialization method \cite{Young-Ma2019xavier}. 
The primary reason for choosing this architecture was related to the language aspects of the problem, where SQL queries are treated as sentences with a structured order of clauses and tokens. 
An LSTM network has proved its efficiency in learning complex sequential data, which was our initial motivation for selecting this architecture~\cite{lipton2015critical}.
Upon data updates, we retrain the LSTM from its last state on a training set consists of queries which span on the new data.

%% file: Sections/eval.tex
\section{\label{sec:Evauation}Evaluation}
\subsection{\label{subsec:datasets}Datasets}
The system was evaluated using 12 unique data sets, both proprietary and open source. 
The data sets characteristics are presented in Table~\ref{table:data_chars}.

\subsection{\label{subsec:datapartition}Training set partitioning}

For each dataset, a training set was generated and split, using python package sklearn \.cross\_validation (\.train\_test\_split), to three datasets: (1) training set - 70\% of the queries, (2) validation set - 15\% and (3) testing set - 15\%. 

\subsection{\label{subsec:lstmcostfunction}LSTM training cost function}

The LSTM network is trained to minimize a quadratic cost function, defined as:

\begin{equation}\label{MSE}
MSE=\frac{1}{n}\sum^n_{i=1}{{\left({\hat{Y}}_i-\ Y_i\right)}^2}\
\end{equation}

\noindent where $Y_i$ is the real query result, ${\hat{Y}}_i$ is the model approximated query result, and $n$ is the batch size.
\begin{table*}[h]
\tabcolsep=0.008cm
\small
\caption{Datasets characteristics.}
\label{table:data_chars}
\begin{tabular}{|c|c|c|c|c|c|c|c|c|c|c|}
\hline
\textbf{} & \textbf{Dataset} & 
\textbf{\begin{tabular}[c]{@{}c@{}}Proprietary\\  data source\end{tabular}} & 
\textbf{\begin{tabular}[c]{@{}c@{}}Target \\ function\end{tabular}}   & \textbf{\begin{tabular}[c]{@{}c@{}}\# \\ $attr^{(n)}$\end{tabular}} & \textbf{\begin{tabular}[c]{@{}c@{}}\# \\$attr^{(c)}$\end{tabular}} & \textbf{\begin{tabular}[c]{@{}c@{}}\# rows\end{tabular}} & \textbf{\begin{tabular}[c]{@{}c@{}}\# \\ queries\end{tabular}} & \textbf{\begin{tabular}[c]{@{}c@{}} Mean \\Entropy\end{tabular}} & \textbf{\begin{tabular}[c]{@{}c@{}}Input\\Tensor\\   variance\end{tabular}} & \textbf{\begin{tabular}[c]{@{}c@{}}Target\\  column STD\end{tabular}} \\ \hline
1 & average\_revenue & Yes & \begin{tabular}[c]{@{}c@{}}avg\\(revenue)\end{tabular} & 3 & 2 & 1000000000 & 5205078                                                           & 6.293                                                                   & 0.154                                                                & 40400000                                                                  \\ \hline
2           & \begin{tabular}[c]{@{}c@{}}average\_success\\ \_rate\end{tabular}         & Yes                                                                                & \begin{tabular}[c]{@{}c@{}}avg\\ (build\_time)\end{tabular}           & 2                                                           & 3                                                          & 2333293                                                              & 415791                                                            & 2.264                                                                   & 5.421                                                                & 19.196                                                                    \\ \hline
3           & \begin{tabular}[c]{@{}c@{}}count\_product\\ \_pass\end{tabular}         & Yes                                                                                & \begin{tabular}[c]{@{}c@{}}count\\ (machine\_id)\end{tabular}         & 1                                                           & 5                                                          & 4000000000                                                           & 811928                                                            & 0.942                                                                   & 0.151                                                                & 2350516                                                                   \\ \hline
4           & \begin{tabular}[c]{@{}c@{}}count\_product\\ \_fail\end{tabular}         & Yes                                                                                & \begin{tabular}[c]{@{}c@{}}count\\ (machine\_id)\end{tabular}         & 1                                                           & 5                                                          & 95484                                                                & 451173                                                            & 0.942                                                                   & 0.153                                                                & 8613                                                                      \\ \hline
5           & \begin{tabular}[c]{@{}c@{}}count\_product\_\\ false\_calls\end{tabular} & Yes                                                                                & \begin{tabular}[c]{@{}c@{}}count\\ (machine\_id)\end{tabular}         & 1                                                           & 5                                                          & 350232                                                               & 378111                                                            & 0.942                                                                   & 0.202                                                                & 215315                                                                    \\ \hline
6           & \begin{tabular}[c]{@{}c@{}}count\_churn\_\\ customers\end{tabular}      & Yes                                                                                & \begin{tabular}[c]{@{}c@{}}count\\ (customer\_id)\end{tabular}        & 4                                                           & 3                                                          & 9263836                                                              & 62092                                                             & 2.782                                                                   & 0.13267                                                              & 530                                                                  \\ \hline
7           & \begin{tabular}[c]{@{}c@{}}sum\_duration\_\\ call\end{tabular}          & Yes                                                                                & \begin{tabular}[c]{@{}c@{}}sum\\ (duration)\end{tabular}              & 3                                                           & 2                                                          & 9349                                                                 & 100000                                                            & 3.198                                                                   & 0.167                                                                & 861                                                                       \\ \hline
8           & \begin{tabular}[c]{@{}c@{}}average\_ibm\\ \_price\end{tabular}          & No                                                                                 & \begin{tabular}[c]{@{}c@{}}avg\\ (close\_price)\end{tabular}          & 1                                                           & 2                                                          & 1048575                                                              & 340489                                                            & 0.343                                                                   & 0.125                                                                & 471                                                                       \\ \hline
9           & \begin{tabular}[c]{@{}c@{}}average\_realestate\\ \_price\end{tabular}  & No                                                                                 & \begin{tabular}[c]{@{}c@{}}avg\\ (price)\end{tabular}                 & 3                                                           & 2                                                          & 22489348                                                             & 508086                                                            & 2.113                                                                   & 0.105                                                                & 236                                                                       \\ \hline
10          & \begin{tabular}[c]{@{}c@{}}avg\_stock\_close\\ \_price\end{tabular}     & No                                                                                 & \begin{tabular}[c]{@{}c@{}}avg\\ (close\_price)\end{tabular}          & 2                                                           & 1                                                          & 63267                                                                & 8721                                                              & 5.703                                                                   & 0.157                                                                & 118                                                                       \\ \hline
11          & \begin{tabular}[c]{@{}c@{}}average\_paid\\ \_days\end{tabular}          & Yes                                                                                & \begin{tabular}[c]{@{}c@{}}avg\\ (actual\_paid\\ \_days)\end{tabular} & 3                                                           & 2                                                          & 100000000                                                            & 508365                                                            & 0.451                                                                   & 0.099                                                                & 25667553                                                                     \\ \hline
12          & \begin{tabular}[c]{@{}c@{}}average\_build\_\\ duration\end{tabular}     & Yes                                                                                & avg(duration)                                                         & 1                                                           & 3                                                          & 22276094                                                             & 325935                                                            & 0.993                                                                   & 0.129                                                                & 7487                                                                      \\ \hline
\end{tabular}
\end{table*}

\subsection{\label{sec:evaluationmetrics}Evaluation metrics}

Since the target variable (query result) is continuous, a simple regression cost function such as MSE or RMSE can yield an unnormalized range of values and is greatly influenced by the problem scale. \\
For this reason, we have chosen a normalized version of RMSE (NRMSE), defined as follows:
\noindent 
\begin{equation}\label{RMSE}
RMSE=\sqrt{\frac{\sum^n_{i=1}{{({\hat{Y}}_i-\ Y_i)}^2}}{n}}\
\end{equation}
\noindent where $i$ represent a query from testing set, $Y_i$ is the real query result, ${\hat{Y}}_i$ is the model approximated query result, and $n$ is number of testing set queries.

\begin{equation}\label{NRMSE}
NRMSE=\frac{RMSE}{Y_{max}-\ Y_{min}}    
\end{equation}

\noindent where $Y_{max}$ is the max query result and $Y_{min}$ is the min query result. 
Normalizing the RMSE facilitates the comparison between data sets with different scales. \\
In addition, we calculated $QL$ -- the duration of a single query execution in milliseconds, as well as the queries throughput at batch mode (using GPU), referred to as $QT$ 
 and measured as follows:
\begin{equation}\label{pr}
QT=\frac{T}{Q}    
\end{equation}
where $\mathrm{\ }T$ is the total latency of the batch mode prediction operation and $Q$ is the number of queries used in the testing set.\\
\\
Lastly we calculate $ME$ mean entropy $\overline{\mathrm{H}}$ for data set predictors (all columns in \textit{where} clause), while for each categorical column entropy is calculated as:
\[\mathrm{H}\left(\mathrm{X}\right)\mathrm{=\ -}\sum^n_{i\mathrm{=1}}{P\mathrm{(}X_i}\mathrm{)}{Log}^{P\mathrm{(}X_i\mathrm{)}}_{\mathrm{2}}\] 
Where
\\
$\mathrm{\ }i$ varies from 1 to n -- number of distinct values for a categorical column and

\noindent ${P\mathrm{(}X}_i\mathrm{)}$ -- is the number of rows containing value $i$ divided by total number of rows. For continuous columns, we first discretize them to categorical columns using 10 equal-length bins and then use the above entropy calculation. We chose 10 bins to allow enough bins to capture the column variance, but not so many as to prevent reasonable calculation time.
\[\overline{\mathrm{H}}=\ \frac{\sum^M_{j=1}{{H(X)}_j}}{M}\] 
Finally, we calculated the LSTM input layer variance 
using Tensorflow moments \footnote{\url{https://www.tensorflow.org/api_docs/python/tf/nn/moments}}.

Data complexity which is expressed by the data mean entropy $ME$, target column STD (standard deviation) determine the problem complexity and  influences how well and how quick the LSTM can learn the data converge. 
According to these measures, we have configured a hyper-parameters heuristic that we found, by trial and error, to be effective. 
\subsection{\label{subsec:dnncostfunction}DNN training cost function}

The DNN is trained to minimize a quadratic cost function, which is also known as mean squared error, maximum likelihood, and sum squared error, defined as:
\begin{equation}\label{eq:MSE}
MSE=\frac{1}{n}\sum^n_{i-1}{{\left({\hat{Y}}_i-\ Y_i\right)}^2}\
\end{equation}
\noindent $Y_i$ -- real query result

\noindent ${\hat{Y}}_i$ -- model approximated query result

\noindent $n$ -- batch size\\


\noindent \textbf{Baseline methods and metrics.}
We have chose three baseline methods: (1) VerdictDB \cite{park2018verdictdb}, a novel AQP method  which accelerates analytical queries and (2) BlinkDB \cite{agarwal2013blinkdb} - an approximate query engine for running interactive SQL queries on large volumes of data.
We use mean query latency (QL) and NRMSE metrics to evaluate performance and accuracy respectively.


\subsection{\label{sec:experimentresults}Experiment Results}
For each data set, Table~\ref{table:models_performance} specifies LSTM network training parameters and the trained model performance metrics. \\

\begin{table*}[h]
\small
\centering
\caption{Datasets models performance and accuracy metrics.}
\label{table:models_performance}
\tabcolsep=0.008cm
\begin{tabular}{|c|c|c|c|c|c|c|c|c|c|c|c|} 
\hline
\textbf{}    & \multicolumn{7}{c|}{\textbf{LSTM architecture and hyper-parameters} }                                                                                                                                                                                                                                                                                                & \multicolumn{4}{c|}{\textbf{Model performance} }                                                                                                                                                                                                \\ 
\hline
\textbf{\#}  & \textbf{Dataset name}                                                 & \textbf{LR}  & \textbf{Batch size}  & \begin{tabular}[c]{@{}c@{}}\textbf{LSTM}\\\textbf{ neurons} \end{tabular} & \begin{tabular}[c]{@{}c@{}}\textbf{Dense}\\\textbf{ Neurons} \end{tabular} & \begin{tabular}[c]{@{}c@{}}\textbf{input}\\\textbf{ shape} \end{tabular} & \textbf{GPU Type}  & \begin{tabular}[c]{@{}c@{}}\textbf{QT}\\\textbf{ (q/s)} \end{tabular} & \begin{tabular}[c]{@{}c@{}}\textbf{QL}\\\textbf{ (ms/q)} \end{tabular} & \textbf{NRMSE}  & \begin{tabular}[c]{@{}c@{}}\textbf{LSTM}\\\textbf{ size (Mb)} \end{tabular}  \\ 
\hline
1            & average\_revenue                                                       & 1.E-05       & 2048                 & 512                                                                       & 200                                                                        & (7,17)                                                                   & GTX 2060               & 25501                                                                 & 2.79                                                                   & 1.57            & 3.51                                                                         \\ 
\hline
2            & average\_success\_rate                                                  & 1.E-02       & 1024                 & 128                                                                       & 200                                                                        & (7,17)                                                                   & GTX  2060              & 43512                                                                 & 2.55                                                                   & 3.75            & 2.51                                                                         \\ 
\hline
3            & count\_product\_pass                                                    & 1.E-04       & 2048                 & 512                                                                       & 400                                                                        & (16,18)                                                                  & GTX 2060               & 1881                                                                  & 27.41                                                                  & 0.07            & 3.84                                                                         \\ 
\hline
4            & count\_product\_fail                                                    & 1.E-03       & 1024                 & 512                                                                       & 400                                                                        & (16,18)                                                                  & AWS K80                & 6185                                                                  & 28.60                                                                  & 0.51            & 3.84                                                                         \\ 
\hline
5            & \begin{tabular}[c]{@{}c@{}}count\_product\_false\\ \_calls \end{tabular} & 1.E-04       & 1024                 & 512                                                                       & 400                                                                        & (16,18)                                                                  & AWS K80                & 1889                                                                  & 28.2                                                                   & 0.12            & 3.84                                                                         \\ 
\hline
6            & \begin{tabular}[c]{@{}c@{}}count\_churn\_\\ customers \end{tabular}    & 1.E-02       & 2048                 & 128                                                                       & 200                                                                        & (17,17)                                                                  & AWS K80                & 21304                                                                 & 3.55                                                                   & 0.14            & 2.56                                                                         \\ 
\hline
7            & sum\_duration\_all                                                     & 1.E-02       & 128                  & 256                                                                       & 200                                                                        & (7,20)                                                                   & GTX 2060                & 1877                                                                  & 2.84                                                                   & 0.01            & 2.58                                                                         \\ 
\hline
8            & average\_ibm\_price                                                     & 1.E-02       & 128                  & 256                                                                       & 200                                                                        & (7,62)                                                                   & AWS K80                & 26667                                                                 & 3.29                                                                   & 0.61            & 2.91                                                                         \\ 
\hline
9            & \begin{tabular}[c]{@{}c@{}}average\_realestate\\ \_price \end{tabular}  & 1.E-02       & 1024                 & 128                                                                       & 200                                                                        & (13,61)                                                                  & GTX 2060                 & 76086                                                                 & 2.81                                                                   & 0.32            & 2.9                                                                          \\ 
\hline
10           & \begin{tabular}[c]{@{}c@{}}avg\_stock\_close\\ \_price \end{tabular}     & 1.E-02       & 2048                 & 128                                                                       & 200                                                                        & (6,61)                                                                   & GTX  2060              & 121259                                                                & 4.03                                                                   & 0.17            & 2.9                                                                          \\ 
\hline
11           & average\_paid\_days                                                     & 1.E-05       & 2048                 & 128                                                                       & 200                                                                        & (27,7)                                                                   & AWS k80                & 62046                                                                 & 2.05                                                                   & 0.97            & 2.63                                                                         \\ 
\hline
12           & \begin{tabular}[c]{@{}c@{}}average\_build\\ \_duration \end{tabular}    & 1.E-04       & 1024                 & 256                                                                       & 200                                                                        & (10,32)                                                                  & GTX 2060               & 12159                                                                 & 4.01                                                                   & 0.15            & 2.74                                                                         \\
\hline
\end{tabular}
\end{table*}

\noindent \textbf{Accuracy.}
As expected, NRMSE for the largest model (dataset \#5)  with 512 LSTM neurons layer and a dense layer with 400 neurons, was the smallest (most accurate) with a value of 0.12 while for the smallest model (dataset \#6) with 128 LSTM neurons layer and a dense layer with 200 neurons, was the largest with a value of 3.75.
\noindent \textbf
{Query latency performance.}
Using GPU, a throughput (QT) of approximately 121k queries per second was measured, while a single query latency (QL) for our largest (slowest) model lasted 28 ms.
Generally and as can be expected, as the LSTM network is more complex (more layers, more neurons, larger input), latency goes up and throughput goes down.\\

\begin{figure}[htb]
  \centering
  \includegraphics[scale=0.30]{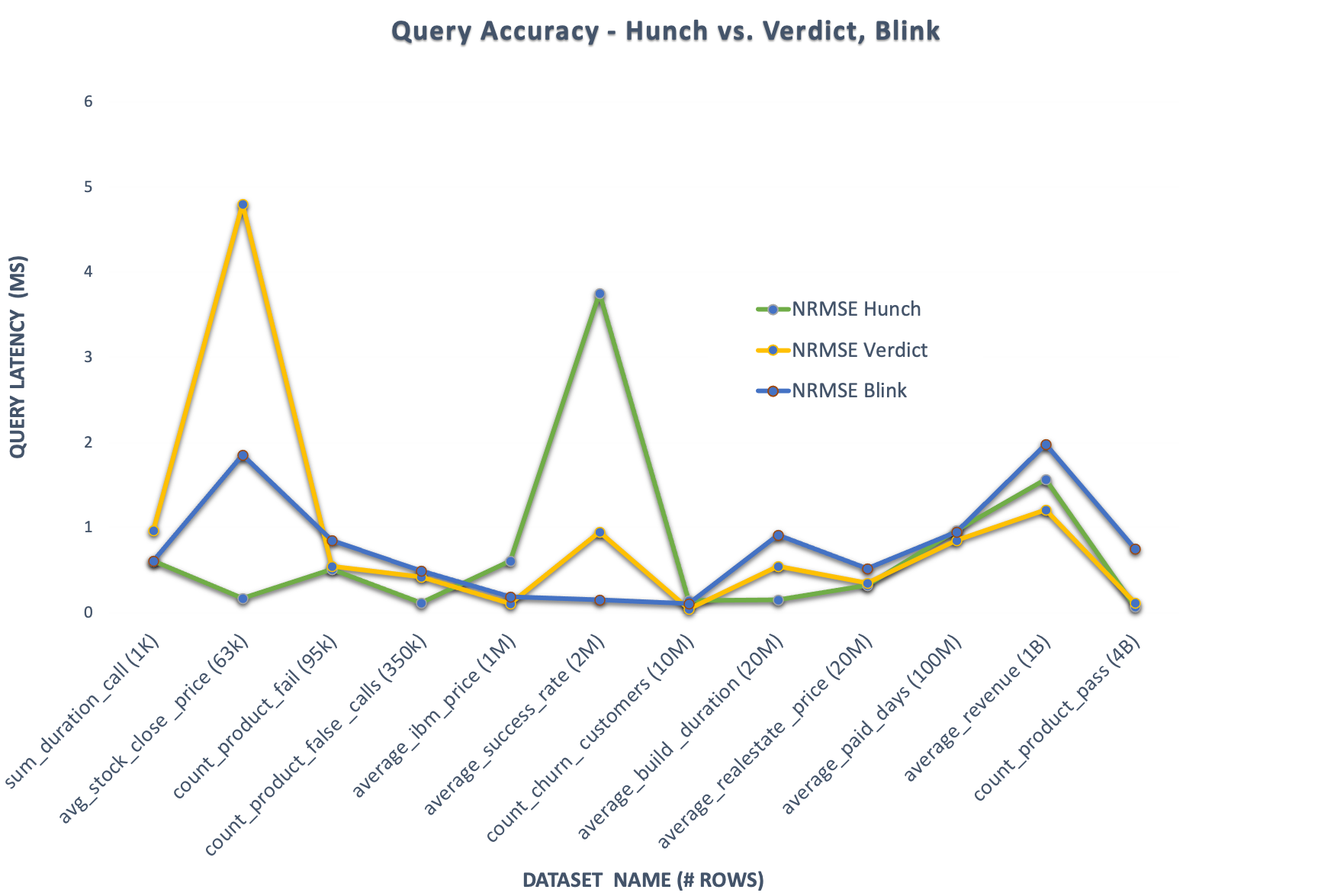}
  \caption{Comparing the accuracy (NRMSE) performance of Hunch Verdict and Blink).}
  \label{fig:compare_accur}
  \vspace{6pt}
  \includegraphics[scale=0.30]{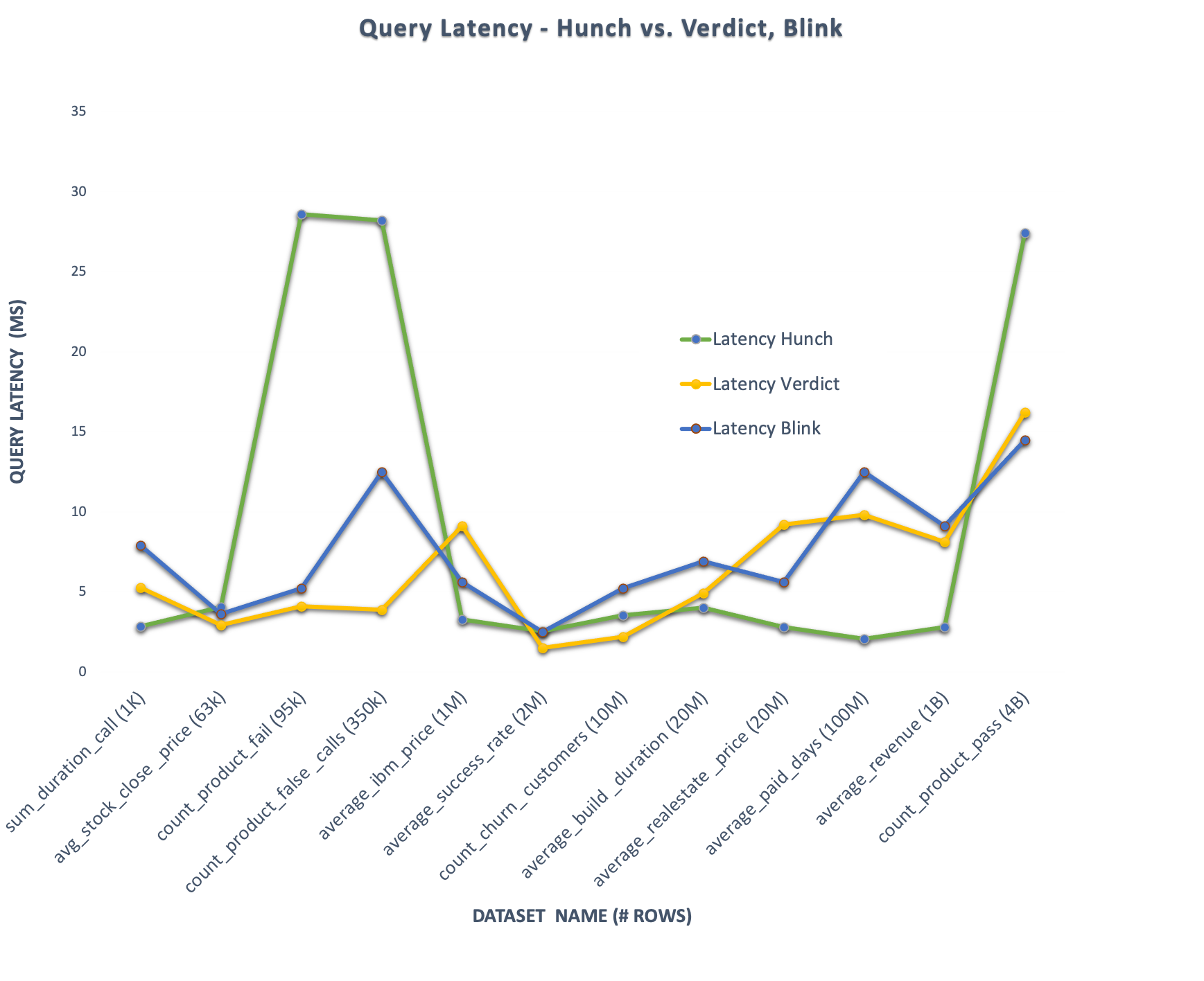}
  \caption{Comparing the latency performance of Hunch with Verdict and Blink).}
  \label{fig:compare_laten}
  \centering
\end{figure}

\begin{figure}[htb]
  \centering
  \includegraphics[scale=0.37]{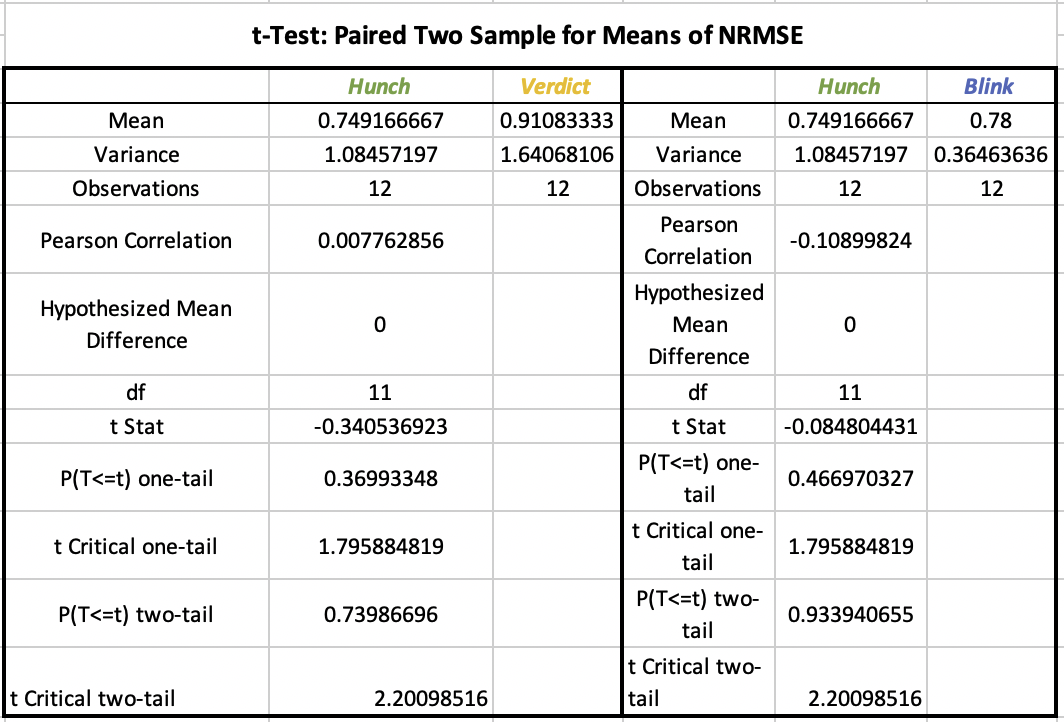}
  \caption{T-test analysis comparing Hunch with Verdict and Blink on NRMSE
  \label{fig:compare_laten_paired}}
  \vspace {8mm}
  \includegraphics[scale=0.37]{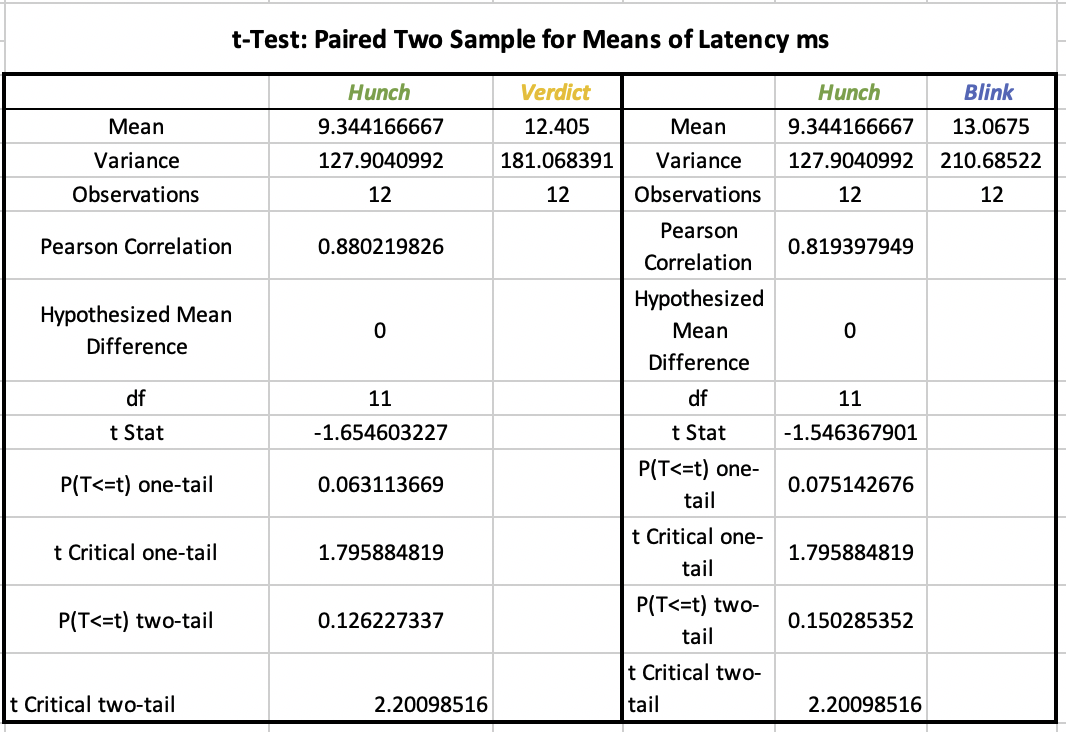}
  \label{fig:compare_acc_paired}
   \caption{T-test analysis comparing Hunch with Verdict and Blink latency metric.}
   
  \centering
\end{figure}

\noindent\textbf{Baseline comparison results.}
Figure~\ref{fig:compare_accur} and Figure~\ref{fig:compare_laten} depicts the accuracy and latency of Hunch and baseline methods - VerdictDB~\cite{park2018verdictdb}, and BlinkDB \cite{agarwal2013blinkdb} on all datasets. 
Figures~\ref{fig:compare_acc_paired}  show a paired T-test analysis to determine whether our method were found statistically better than the compared methods in terms of accuracy (NRMSE) and query latency. From these figure, although it is evident that our method was better the majority of data sets (both on the accuracy and latency plots), statistically one cannot claim one method is superior to the other. 


%% file: Sections/discussion.tex
\subsection{\label{sec:Discussion}Discussion}

Our proposed solution predicts query results within a controlled accuracy (NRMSE), ranging between approximately 1\% to 4\%.
QL ranges from approximately 2.0 ms/q to 30.0 ms/q for a single query. 
Moreover, for large datasets (20M - 4B records), our method is two orders of magnitude faster from the compared methods.
In batch mode (utilizing GPU batch processing) the method was capable of calculating results at up to 120,000 queries per second. 
These encouraging results led us to consider this solution as a novel query approximation tool, capable of saving heavy-lifting database processing and data transfer from the consumer to the database and back. 
Our method can predict missing data points and data points that span in the future. 
For instance, when the system was trained on temporal dimensions (e.g., dates in \textit{where} clause), our 10${}^{th}$ dataset (named avg\_stock\_close in Table~\ref{table:data_chars})
results shows a NRMSE of approximately 0.2\% for testing set with future dates (which were not available during training). 
Comparing our method to other state-of-the-art AQP methods, our method is resilient to growing scale of datasets (2M rows). 
This is because our method uses the LSTM network to calculate SQL query's result and is decoupled from the dataset after training process is done.

Although statistically we have not established supremacy of our method over the compared methods in terms of accuracy and latency, we do believe our method has the advantage of being lean (3Mb in average), thus enabling light and rapid deployment on client production sites and end devices such as cellular phones and tablets. This could be particularly useful since once our model is deployed on such device, it can run queries on very large data sets without relying on internet connection.

The shortcomings of our method are as follow : (1) the limited query structure model which currently does not support join, exist operation and other sub-queries operations. This limitation can be mitigated by persisting a join query as a prepossessing step. (2) Changes in data sets. When new data is appended, it requires retraining the LSTM from the last training checkpoint (network state) with queries that are relevant to new data records. The operation of retraining the LSTM requires significantly lower efforts and is speculated to last only a fraction of the first LSTM training time.

\section{\label{sec:futurework}Conclusion and Future Research}

The primary goal of this research was to develop a novel approach for AQP over massive datasets.
A secondary goal was to show how our method can serve as an interactive analytical tool, demonstrating rapid responses processing a queries on large datasets. 
Existing querying methods require ongoing access to the underlying data. 
Once training is done, our proposed method does not require online connection to the data to approximate SQL queries. 
This opens up an array of potential use cases for big data analytics.
Going forward, we aim to develop our method to scenarios where dataset changes frequently, thus the model should adapt more quickly to new data. 
In addition, we plan to enrich the method to support additional SQL operations.
\footnote{Sisense®, and Hunch™ are trademarks of Sisense Ltd. All other trademarks are property of their respective owners. The Hunch Platform and additional products not listed here are covered by US and foreign patents. Additional patent applications may be pending. 
}
\section*{Acknowledgement}
We want to thank Dr. Guy Levy-Yurista and Mr. Adi Azaria for supporting and mentoring this project as their management, encouragement, suggestions and comments were insightful and invaluable. We also want to thanks Sisense Ltd. for hosting and supporting this research and all the required resources.